\documentclass[prb,superscriptaddress,twocolumn,floatfix,letterpaper]{revtex4}
\usepackage{amssymb, amsmath}
\usepackage{mathrsfs}
\usepackage[usenames]{color}
\usepackage{subfigure, wrapfig}
\usepackage[dvips]{graphicx}
\usepackage[normalem]{ulem}

\newcommand{\txt}[1]{\textnormal{#1}}
\newcommand{\punct}[1]{\textrm{ #1}}
\newcommand{\etal}{{\itshape et al.\@ }}
\ifx\printedBW\undefined
	\newcommand{\colorfig}{}
\else
	\newcommand{\colorfig}{(Color online) }
\fi

\newcommand{\vv}[1]{\mathbf{#1}}

\newcommand{\dg}{\dagger}
\newcommand{\up}{\uparrow}
\newcommand{\dn}{\downarrow}

\newcommand{\avg}[1]{\langle #1 \rangle}
\newcommand{\op}[1]{#1}
\newcommand{\opdag}[1]{#1^\dg}

\newcommand{\fourOps}[5]{\opdag{#1}_{#2} \opdag{#1}_{#3} \op{#1}_{#4} \op{#1}_{#5}}

\newcommand{\JJ}{\mathcal{J}}
\newcommand{\EE}{\mathcal{E}}
\newcommand{\NN}{\mathcal{N}}
\newcommand{\Uc}{U_c^{\txt{Mott}}}

\usepackage[dvips]{hyperref}

\begin{document}

\title{Magnetism and Mott Transition: A Slave-rotor Study}

\author{Wing-Ho Ko}
\affiliation{Kavli Institute for Theoretical Physics, University of California, Santa Barbara, Santa Barbara, California, 93106, USA}
\author{Patrick A. Lee}
\affiliation{Department of Physics, Massachusetts Institute of Technology, Cambridge, Massachusetts, 02139, USA}

\date{\today}

\begin{abstract} 
Motivated by the debate of spin-density-wave (SDW) versus local-moment (LM) picture in the iron-based superconducting (FeSC) materials, we consider a two-band orbital-symmetric Hubbard model in which there is robust Fermi surface nesting at $(\pi,0)$. We obtain the phase diagram of such system by a mean-field slave-rotor approach, in which the Fermi surface nesting and the SDW order are explicitly taken into account via a natural separation of scale between the Hund's coupling and the Coulomb interaction. We find that for a sizable range of Hund's coupling the Mott transition acquires a strong first-order character, but there also exists a small range of stronger Hund's coupling in which an enhancement of magnetization can be observed on the SDW side. We interpret the former scenario as one in which a sharp distinction can be drawn between LM and the SDW picture, and the latter scenario as one in which signs of LM physics begin to develop in the metallic phase. It is tempting to suggest that some FeSC materials are in the vicinity of the latter scenario.

\end{abstract}

\maketitle 

\section{Introduction} \label{sect:intro} 

Mott transitions in multiband scenarios have recently received increased attention in the condensed matter community, partly owing to the interests in materials related to the iron-based superconducting (FeSC) materials, whose parent compounds are believed to be in close proximity to Mott transitions.\cite{Haule:PRL:2008,Laad:PRB:2009} Indeed, while most parent compounds of the FeSCs are poor metals, insulating behaviors have been observed in\cite{Fang:arXiv:1012.5236} (Tl,K)Fe$_x$Se$_2$ and in\cite{Zhu:PRL:2010} La$_2$O$_2$Fe$_2$O(Se,S)$_2$. Given the sizable magnetic moment of approximately 0.8--1.0~$\mu_B$ in the parent compounds of the ``122'' family FeSCs,\cite{Goldman:PRB:2008, JZhao:PRB:2008b, Jesche:PRB:2008, QHuang:PRL:2008, Kaneko:PRB:2008, Su:PRB:2009} a persistent debate in the field has been whether the magnetism is best described by an itinerant spin-density-wave (SDW) nesting\cite{Mazin:PRL:2008,WLYang:PRB:2009} or a local-moment (LM)\cite{JWu:PRL:2008,Si:PRL:2008} picture. Theoretically, {\itshape ab initio\/} LSDA calculation\cite{Boeri:PRB:2010} is able to obtain the experimentally observed magnetic ordering pattern in these compounds, but the magnetic moment is overestimated to be approximately 2.0~$\mu_B$, suggesting that a weak-coupling SDW picture alone may not be adequate in describing these materials.

With the above considerations, it is beneficial to consider how magnetism and Fermi surface nesting affect the Mott transition, and vice versa, in a multiband scenario. Of particular interest is the question of whether there are any signs of local-moment physics on the metallic side of the phase diagram. However, while multiband Mott transitions have previously been studied via dynamical mean-field theory \cite{Werner:PRL:2007, Werner:PRB:2009,Ishida:PRB:2010} and slave-spin mean-field,\cite{Yu:arXiv:1006.2337} many of these studies have been focused on paramagnetic states. 

In contrast, in this paper we present a slave-rotor\cite{Florens:PRB:2004} study of an orbital-symmetric multiband Mott transition in which, utilizing a natural separation of scale between the orbital-symmetric Coulomb repulsion and the Hund's coupling, the SDW Fermi surface nesting has been explicitly taken into account. We find that for a sizable range of Hund's coupling, the existence of (nearly) nesting Fermi surfaces causes the Mott transition to acquire a strong first-order character, in which the (staggered) magnetization jumps across the phase boundary. However, as the strength of Hund's coupling further increases, the first-order transition becomes weaker and an enhancement of magnetization can be observed on the metallic side. We interpret the former scenario as one in which a sharp distinction can be drawn between the itinerant SDW nesting and the LM picture, and the latter scenario as one in which signs of LM physics begin to develop in the metallic phase.  

It is worth noting that in the region of parameter space in which the staggered magnetization is enhanced, the renormalized hopping parameters also develop a discernible anisotropy between the $x$ and $y$ nearest-neighbor bonds, in which the antiferromagnetic $x$ direction shows a larger renormalized hopping than the ferromagnetic $y$ direction. Such anisotropy is insignificant in other regions of the parameter space we considered. In the context of FeSC, such anisotropy may be associated and is in agreement with the observed resistivity anisotropy in detwined Ba(Fe$_{1-x}$Co$_x$)$_2$As$_2$ samples,\cite{Chu:Sci:2010} in which the in-plane resistivity in the antiferromagnetic direction is found to be smaller than that in the ferromagnetic direction---a result difficult to reconcile with a simple Landau Fermi liquid picture.\cite{Laad:arXiv:1010.2940}

We also remark that the renormalized bandwidth we obtained in this region of enhanced magnetization is about 2.5--5 times narrower than that of the corresponding noninteracting tight-binding model---values that are close to, albeit larger, than those reported for some FeSC materials.\cite{Lu:Nat:2008, Terashima:PNAS:2009} Similarly, the staggered magnetization in such region is around 1.3--1.5~$\mu_B$, close to but slightly larger than the values observed in the FeSC materials in the ``122'' family. Together with the observed resistivity anisotropy, it is tempting to suggest that some FeSC materials in the ``122'' family are close to this region of enhanced magnetization that we associated with remnant of LM physics in the metallic phase.

\section{The Multiband Hubbard Model and the Slave-rotor formulation} \label{sect:formulation}

A multiband Mott transition can be described by a multiband Hubbard model, which can be written as the sum of a noninteracting hopping Hamiltonian and an on-site interaction term:
\begin{align}
%\begin{equation}
H & = H_{\txt{hop}} + H_{\txt{int}} \punct{,} \label{eq:H_Hb} \\
%\end{equation}
%\begin{equation}
H_{\txt{hop}} & = \sum_{i,j} \sum_{a,b} (t^{ab}_{ij} \opdag{c}_{ia\sigma} \op{c}_{jb\sigma} + h.c.) - \mu \sum_{i} \sum_{a} \opdag{c}_{ia\sigma} \op{c}_{ia\sigma} \punct{,} \label{eq:H_hop} \\
%\end{equation}
%\begin{align}
H_{\txt{int}} & = 
	\sum_i \Big( U \sum_{a} \fourOps{c}{ia\sigma}{ia\sigma'}{ia\sigma'}{ia\sigma} \notag \\
	& + \frac{U-2 J_1}{2} \sum_{a\neq b} \fourOps{c}{ia\sigma}{ib\sigma'}{ib\sigma'}{ia\sigma} \notag \\
	& + \frac{J_2}{2} \sum_{a\neq b} \fourOps{c}{ia\sigma}{ib\sigma'}{ia\sigma'}{ib\sigma} 
	+ \frac{J_3}{2} \sum_{a\neq b} \fourOps{c}{ia\sigma}{ia\sigma'}{ib\sigma'}{ib\sigma} 
\Big) \punct{,} \label{eq:H_int}
\end{align}
where $i,j$ label lattice sites, $a,b$ label orbitals, and $\sigma,\sigma'$ label spins. Note that the sum over repeated spin indices is assumed in the above equations and will be assumed throughout this paper. Note also that we assume $U$ to be orbital independent, as would be case the when all orbitals are atomic. In that limit, it can also be shown\cite{Castellani:PRB:1978} that $J_1 = J_2 = J_3$, which we shall assume henceforth and thus drop the subscripts. Henceforth we shall also refer to the $U$ term as the Coulomb interaction and the $J$ term as the Hund's coupling, even though the latter includes the contributions from co-hopping and inter-orbital Coulomb interactions.

\begin{figure}[tb]%
\begin{center}%
\subfigure[\label{fig:orbitals}]{\includegraphics[width=0.45\columnwidth]{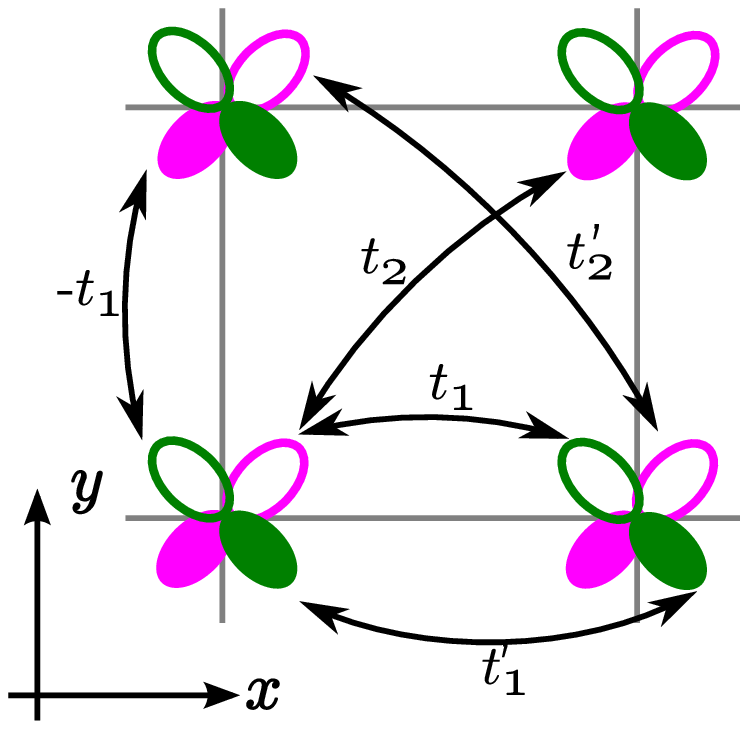}}\quad
\subfigure[\label{fig:FS}]{\includegraphics[width=0.45\columnwidth]{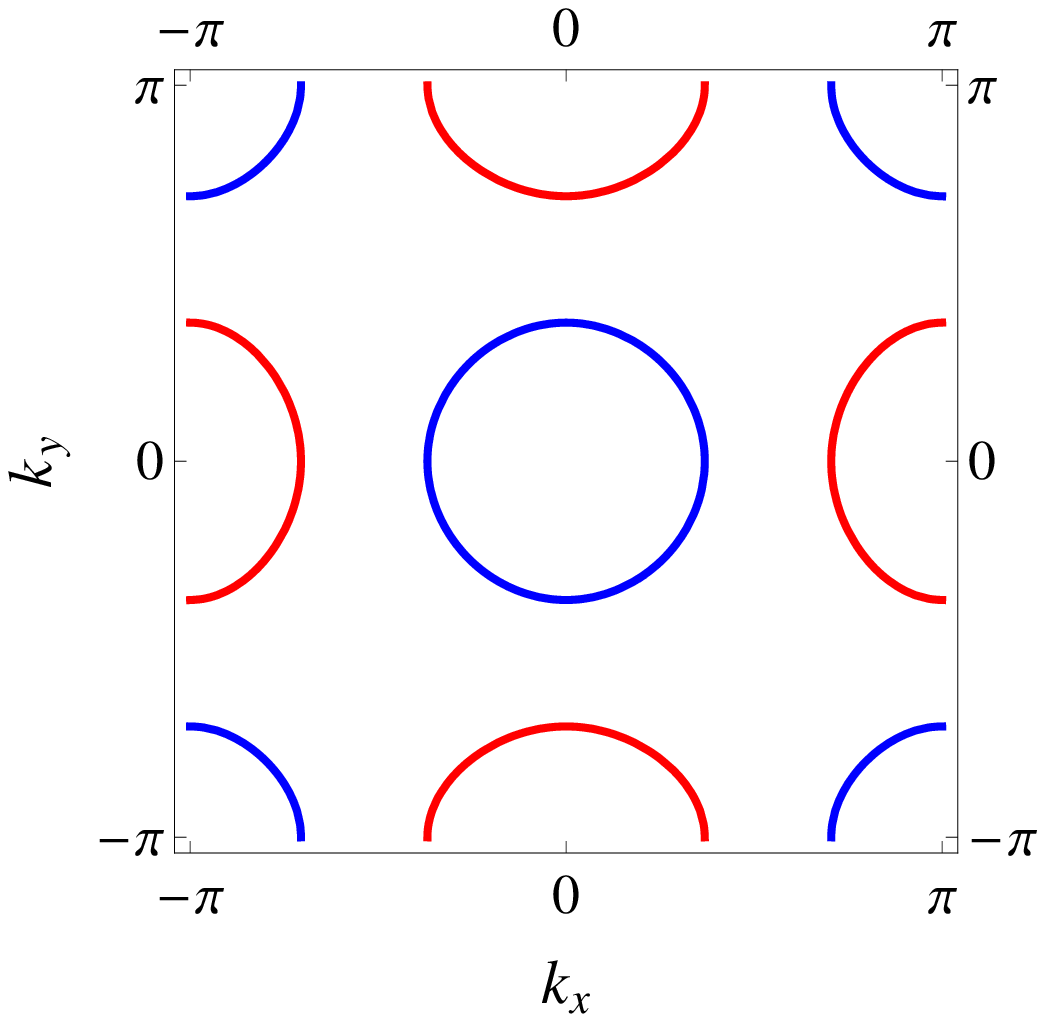}}%
\caption{\label{fig:H_hop}\colorfig (a) The hopping parameters of the two-band hopping Hamiltonian $H_{\txt{hop}}$ in Ref.~\onlinecite{Ran:PRB:2009}; (b) The resulting Fermi surface for $t_1 = 1, t'_1 = 0.2, t_2 = 1.7,$ and $t'_2 = 0.3$ at half-filling (i.e., two electrons per site), where blue represents hole pockets while red represents electron pockets. The choice of parameters and filling fraction in (b) are assumed throughout this paper.}%
\end{center}%
\end{figure}

For concreteness we take $H_{\txt{hop}}$ to be the orbital-symmetric two-band Hamiltonian introduced by Ran \etal in Ref.~\onlinecite{Ran:PRB:2009}, whose hopping parameters and Fermi surface topology are shown in Fig.~\ref{fig:H_hop}. Such Hamiltonian is believed to capture certain essential physics in FeSCs, and in such context the two orbitals can be understood as the $d_{XZ}$ and $d_{YZ}$ orbital of the Fe site, with the $X,Y$ axes oriented in the Fe layer and at 45$^\circ$ with the Fe--Fe bond direction. As is evident from Fig.~\ref{fig:FS}, at half-filling (i.e., two electrons per site) there is strong Fermi surface nesting in $H_{\txt{hop}}$, with nesting vector $(\pi,0)$ and $(0,\pi)$, consistent with the SDW ordering in the\cite{YChen:PRB:2008,Cruz:Nature:2008} ``1111'' and\cite{Goldman:PRB:2008,JZhao:PRB:2008b} ``122'' families FeSCs. However, as explained in Ref.~\onlinecite{Ran:PRB:2009}, for topological reasons the Fermi surfaces are not fully gapped even when interactions are included; hence there is a robust metallic SDW phase in such model.

%H_U & = \sum_i \frac{U}{2} \sum_{a} \fourOps{c}{ia\sigma}{ia\sigma'}{ia\sigma'}{ia\sigma} \label{eq:H_U} \\
%H_J & =	\sum_i \Bigg( \frac{-J_1}{2} \sum_{a\neq b} \fourOps{c}{ia\sigma}{ib\sigma'}{ib\sigma'}{ia\sigma} \notag \\
%	& + \frac{J_2}{2} \sum_{a\neq b} \fourOps{c}{ia\sigma}{ib\sigma'}{ia\sigma'}{ib\sigma} 
%	+ \frac{J_3}{2} \sum_{a\neq b} \fourOps{c}{ia\sigma}{ia\sigma'}{ib\sigma'}{ib\sigma} 
%\Bigg) \punct{,} \label{eq:H_int}

In the slave-rotor formalism,\cite{Florens:PRB:2004} the Hilbert space is enlarged by decomposing the electron operator $\op{c}_{ia\sigma}$ into an $O(2)$ rotor $\theta_i$ and a fermionic spinon $\op{f}_{ia\sigma}$, such that $\op{c}_{ia\sigma} = e^{-i\theta_i} \op{f}_{ia\sigma}$. Note that in such decomposition there is only one rotor per lattice site and the spin and orbital indices are carried solely by $f$. Such economy in the slave-rotor formalism is made possible by the orbital-symmetric structure of the Hubbard Hamiltonian Eq.~\ref{eq:H_Hb}, particularly in the existence of a large orbital-symmetric Coulomb interaction $U$. In this enlarged Hilbert space, the physical subspace is given by the constraint $\sum_a \sum_{\sigma} \left( \opdag{f}_{ia\sigma} \op{f}_{ia\sigma} - 1/2 \right) = L_i$, where $L_i$ is the angular momentum operator conjugate to $\theta_i$.

Plugging the slave-rotor decomposition into the Hubbard Hamiltonian Eq.~\ref{eq:H_Hb}, and applying a mean-field approximation to decouple the spinon $f$ and the rotor $\theta$, we obtain:
\begin{align}
H & \approx H_{MF} = H_f + H_\theta \punct{,} \\
H_f & = \sum_{i,j} \sum_{a,b} (\bar{t}^{ab}_{ij} \opdag{f}_{ia\sigma} \op{f}_{jb\sigma} + h.c.) 
	- \sum_i \sum_{a} (\mu\!+\! h_i) \opdag{f}_{ia\sigma} \op{f}_{ia\sigma} \notag \\
	& + \frac{J}{2} \sum_i \sum_{a\neq b} \Big( -2 \fourOps{f}{ia\sigma}{ib\sigma'}{ib\sigma'}{ia\sigma} \notag \\
	& + \fourOps{f}{ia\sigma}{ib\sigma'}{ia\sigma'}{ib\sigma} 
	+ \fourOps{f}{ia\sigma}{ia\sigma'}{ib\sigma'}{ib\sigma} 
	\Big) \punct{,} \label{eq:H_f} \\
H_\theta & = \sum_{i,j} (\bar{\JJ}_{ij} e^{i (\theta_i - \theta_j)} + h.c.) + \sum_{i} \left( \frac{U}{2} L_i^2 + h_i L_i \right) \punct{,} \label{eq:H_theta}
\end{align}
where $h_i$ are mean-field parameters introduced to enforce the constraint on average,  $\bar{t}^{ab}_{ij} = t^{ab}_{ij} \avg{e^{i(\theta_i-\theta_j)} }$, and $\bar{\JJ}_{ij} = \sum_{a,b} t^{ab}_{ij} \avg{ \opdag{f}_{ia\sigma} \op{f}_{ib\sigma} }$, all of which are to be determined self-consistently. In this paper we consider only translationally invariant solutions in which the rotor unit cell is unenlarged while the spinon unit cell is enlarged to include two lattice sites per cell. Moreover, since the underlying Hamiltonian is time-reversal invariant, we consider only solutions for which both $\bar{\JJ}_{ij}$ and $\bar{t}^{ab}_{ij}$ are real. Furthermore, since the system is at half-filling, the constraint is satisfied by setting $h_i = 0$. In this paper, the self-consistency for $\bar{t}^{ab}_{ij}$ and $\bar{\JJ}_{ij}$ are solved by repeated iterations. 

Following Florens and Georges,\cite{Florens:PRB:2004} a second mean-field approximation is applied to the rotor Hamiltonian Eq.~\ref{eq:H_theta}, in which the rotor variable $e^{i\theta_i}$ is replaced by a complex bosonic field $X_i$ subjected to the constraint $|X_i|^2 = 1$, enforced on average by a second mean-field parameter $\lambda$. With such approximation the rotor Hamiltonian Eq.~\ref{eq:H_theta} reduces to a system of coupled simple harmonic oscillators. The electron quasiparticle weight in this formulation is given by $Z = 1- \int \frac{d^2 \vv{k}}{(2\pi)^2} \sqrt{\frac{U}{4(\EE_X(\vv{k})+\lambda)}}$, where $\EE_X(\vv{k})$ is the dispersion of the $X$ boson in $k$ space. The system is in the metallic phase when $Z \neq 0$, which happens when the $X$ boson condenses at its energy minimum.

For the spinon sector, we apply mean-field factorization to the four-fermion terms that appear in the spinon Hamiltonian Eq.~\ref{eq:H_f}. Specifically, given the Fermi surface nesting, we consider the following SDW mean-field Hamiltonian:
\begin{align}
H_{\txt{SDW}} & = \sum_{i,j} \sum_{a,b} (\bar{t}^{ab}_{ij} \opdag{f}_{ia\sigma} \op{f}_{jb\sigma} + h.c.) - \mu \sum_{i} \sum_{a} \opdag{f}_{ia\sigma} \op{f}_{ia\sigma} \notag \\
	& + \sum_i (-1)^{i_x} \sum_{a,b} M_{ab} \left( \opdag{f}_{ia\up} \op{f}_{ib\up}-\opdag{f}_{ia\dn} \op{f}_{ib\dn} \right)
	\punct{,}
\end{align}
where $i_x$ denotes the $x$ coordinate of site $i$, and $M_{ab}$ is a Hermitian matrix of mean-field parameters, to be determined for given $\bar{t}^{ab}_{ij}$ by minimizing $\avg{H_f}$ with respect to the mean-field state obtained from $H_{\txt{SDW}}$. Note that since $M_{ab}$ carries orbital indices, the possibility of an orbital symmetry broken SDW state is included in our mean-field ansatz.

Since the $H_{\txt{SDW}}$ breaks the symmetry between the $x$ and $y$ direction in the underlying Hamiltonian Eq.~\ref{eq:H_Hb}, the self-consistent mean-field parameters $\bar{\JJ}_{ij}$ and $\bar{t}_{ij}$ can take different values along the $x$ and $y$ direction. We have also allowed for the possibility in which $\bar{\JJ}_{ij}$ and $\bar{t}_{ij}$ take different values along the $(x+y)$ and $(x-y)$ directions, but found no manifestation of such cases in the parameter space we considered.

It is well-noted that different parts of the electron two-body interactions are treated differently in our formulation, with the Coulomb interaction $U$ handled via a ``strong-coupling'' approach while the Hund's coupling $J$ handled via a ``weak-coupling'' approach. While this may seem unsatisfactory at first sight, it may nonetheless capture the qualitative aspects of the system when $U$ is the dominant scale, which seems to hold true for 3d transition metals\cite{Miyake:PRB:2008} and has been assumed in other theoretical calculations involving FeSCs\cite{Daghofer:PRL:2008,FWang:PRL:2009,Ran:PRB:2009,Ishida:PRB:2010}.

\section{Results from Slave-rotor mean field} \label{sect:results}

In this section we present the results of the mean-field calculations outlined in Sec.~\ref{sect:formulation} for the half-filled orbital-symmetric two-band Hubbard Hamiltonian Eq.~\ref{eq:H_Hb}, without worrying about the validity of the approximations made. We shall return to the issue of validity in Sec.~\ref{sect:discussions}. To simplify notations, we measure both $J$ and $U$ in units of $t_1$ (for comparison, note that the full bandwidth of the noninteracting Hamiltonian Eq.~\ref{eq:H_hop} is $W = 12.8~t_1$) and denote $Q^{f}_{ij} = \avg{X_i X^*_j} = \avg{e^{i(\theta_i - \theta_j)}}$, such that $\bar{t}^{ab}_{ij} = Q^{f}_{ij} t^{ab}_{ij}$. We shall refer to $Q^{f}_{ij}$ as the bond renormalization factor, $\bar{t}^{ab}_{ij}$ as the renormalized hopping parameter, and the noninteracting part of $H_f$ in Eq.~\ref{eq:H_f} as the renormalized hopping Hamiltonian. We also denote the orbital-dependent staggered magnetization as $m_{ab}$, i.e., $m_{ab} = \sum_i (-1)^{i_x} \avg{ \opdag{f}_{ia\up}\op{f}_{ib\up} - \opdag{f}_{ia\dn}\op{f}_{ib\dn} } /\NN$ (here $\NN$ is the number of lattice sites), and decompose it as $m_{ab} = \frac{1}{2}(m_0 \delta_{ab} + m_{1} \tau^{1}_{ab} + m_{2} \tau^{2}_{ab} + m_{3} \tau^{3}_{ab}$), with $\tau^{\ell}$ being the Pauli sigma matrices.

\begin{figure}[tb]%
\begin{center}%
\includegraphics[width=\columnwidth]{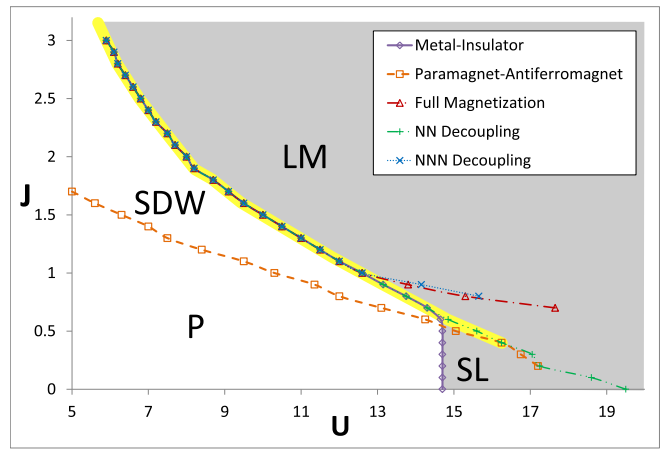}
\caption{\label{fig:phases}\colorfig The phase diagram of the half-filled orbital-symmetric two-band Hubbard Hamiltonian Eq.~\ref{eq:H_Hb}, where $J$ and $U$ are given in units of $t_1$ (for comparison, note that the full bandwidth of the noninteracting system is $W = 12.8~t_1$). Here P stands for paramagnetic metal, SDW stands for spin density wave metal, LM stands for antiferromagnetic insulator, and SL stands for spin liquid. The insulating region is indicated by the gray shade, and first-order phase boundaries are indicated by the yellow shade. For an explanation of the ``NNN decoupling'' and ``NN decoupling'' lines, see Fig.~\ref{fig:decoupling} and the main text.}%
\end{center}%
\end{figure}

\begin{figure}[tb]%
\begin{center}%
\subfigure[\label{fig:J=0.4} $J = 0.4$]{\includegraphics[width=0.95\columnwidth]{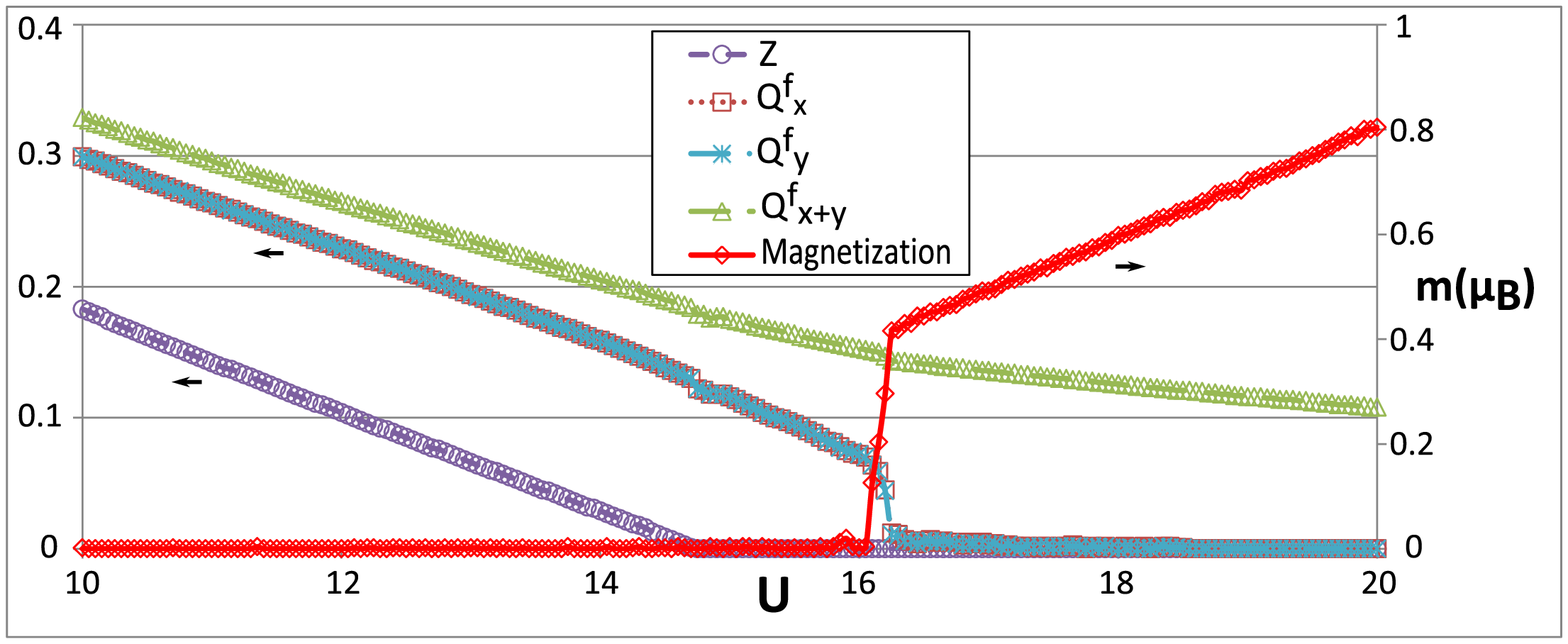}}
\subfigure[\label{fig:J=0.8} $J = 0.8$]{\includegraphics[width=0.95\columnwidth]{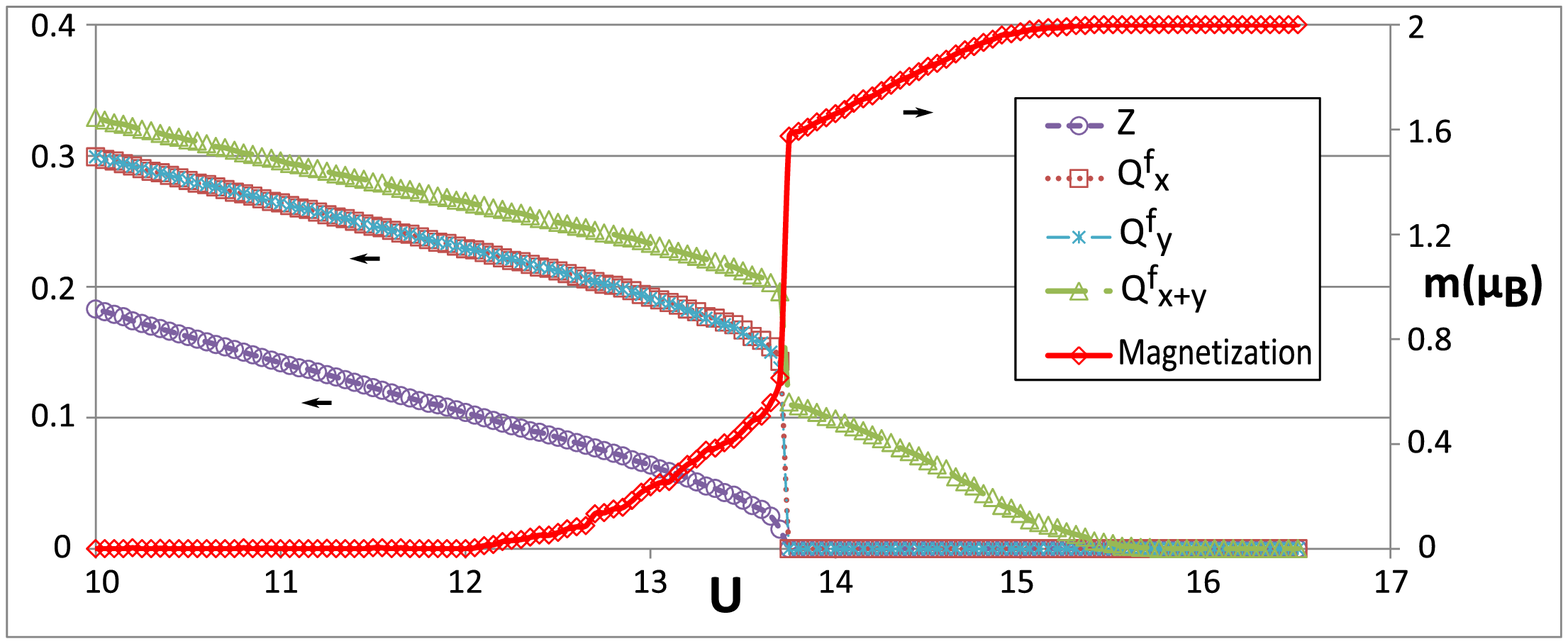}}
\subfigure[\label{fig:J=1.2} $J = 1.2$]{\includegraphics[width=0.95\columnwidth]{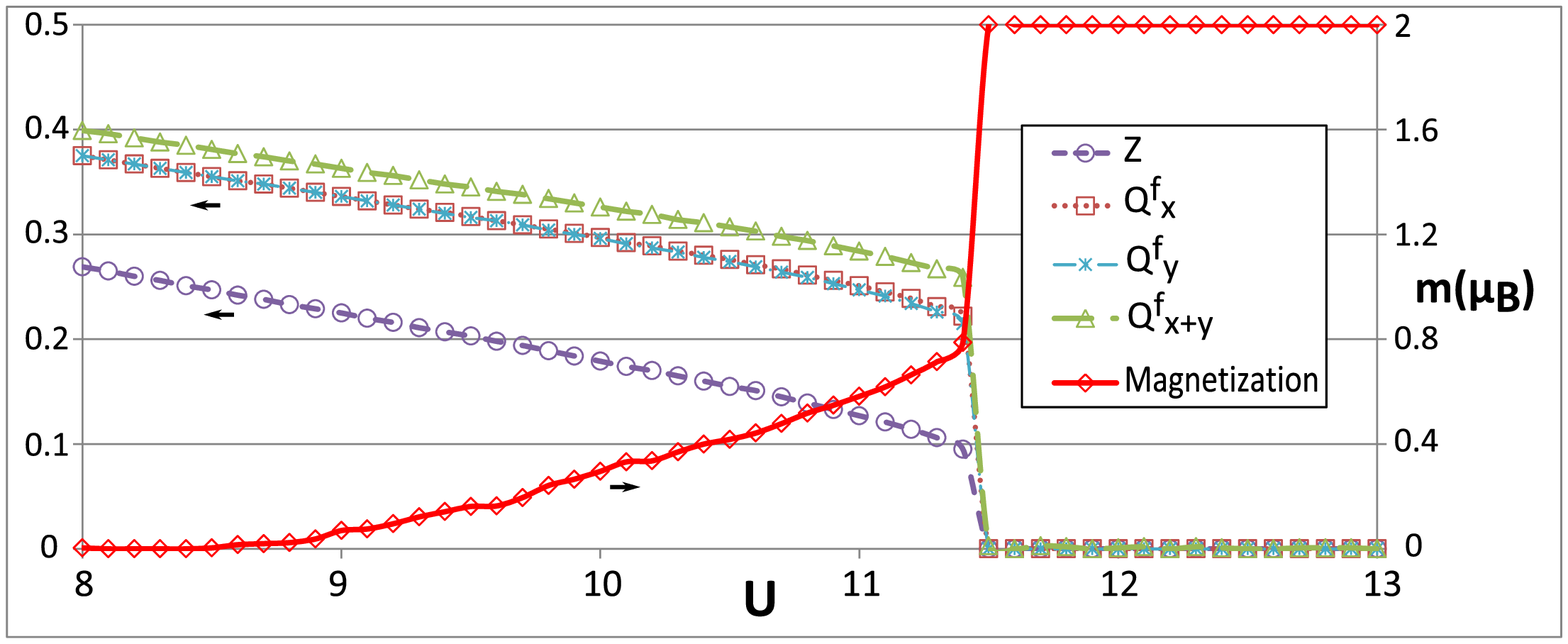}}
\subfigure[\label{fig:J=2.2} $J = 2.2$]{\includegraphics[width=0.95\columnwidth]{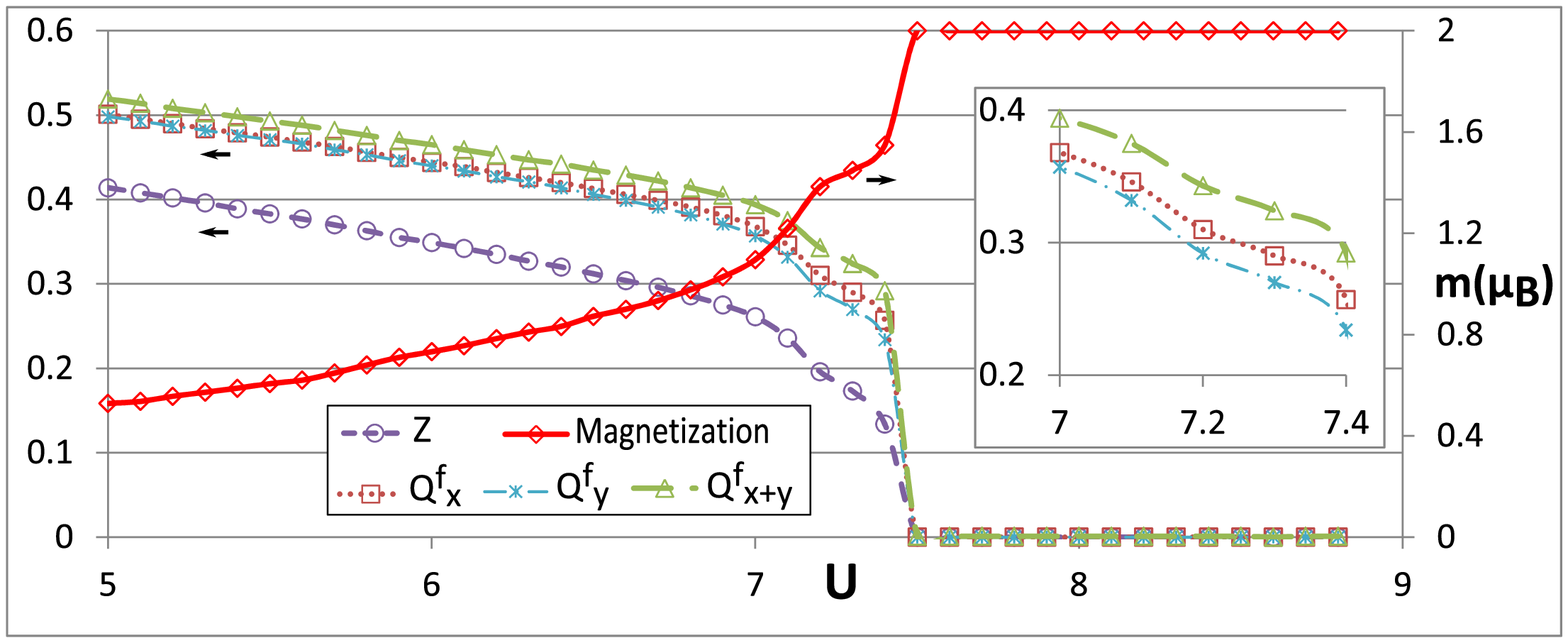}}
\caption{\label{fig:J_cuts}\colorfig The quasiparticle weight $Z$, the bond renormalization factor $Q^{f}_{ij}$ along the $x$, $y$, and diagonal ($x+y$) direction, and the orbital-diagonal staggered magnetization $m_0$ as function of $U$ for (a) $J = 0.4$, (b) $J = 0.8$, (c) $J = 1.2$ and (d) $J = 2.2$. The inset in (d) shows a blowup of the bond renormalization factors in the region where magnetization is enhanced.}%
\end{center}%
\end{figure}

Our major result is the phase diagram shown in Fig.~\ref{fig:phases}, in which we vary $J$ and $U$ while keeping the bare hopping parameters $t^{ab}_{ij}$ fixed. To provide further insight into the phase diagram, we also plot in Fig.~\ref{fig:J_cuts} the quasiparticle weight $Z$, the bond renormalization factor $Q^{f}_{ij}$ along the $x$, $y$, and diagonal ($x+y$) direction, and the orbital-diagonal staggered magnetization $m_0$ as function of $U$ for four characteristic values of $J$. It should be noted that within the parameter range we considered, the orbital-dependent staggered magnetization $m_{ab}$ is always dominated by the orbital-diagonal component $m_0$, in agreement with the weak-coupling results in Ref.~\onlinecite{Ran:PRB:2009}. In other words, no tendency towards an orbital-selective SDW state is found in our results. Because of this, henceforth we shall consider exclusively $m_0$, and refer to it simply as the staggered magnetization.

As typified by Fig.~\ref{fig:J=0.4}, for $J \lesssim 0.5$ we find that the system undergoes a second-order Mott transition before developing any magnetism. The character of the Mott transition is essentially the same as the one considered by Florens and Georges in Ref.~\onlinecite{Florens:PRB:2004}. As a result, for small values of $J$ part of the parameter space supports a non-magnetic insulating state, i.e., a spin liquid. Interestingly, the onset of magnetism for this parameter range seems to coincide with a phase boundary (labeled as ``NN decoupling'' in Fig.~\ref{fig:phases}) in which the nearest-neighbor bonds renormalize to zero. 

\begin{figure}[tb]%
\begin{center}%
\subfigure[\label{fig:NN+NNN}]{\includegraphics[width=0.28\columnwidth]{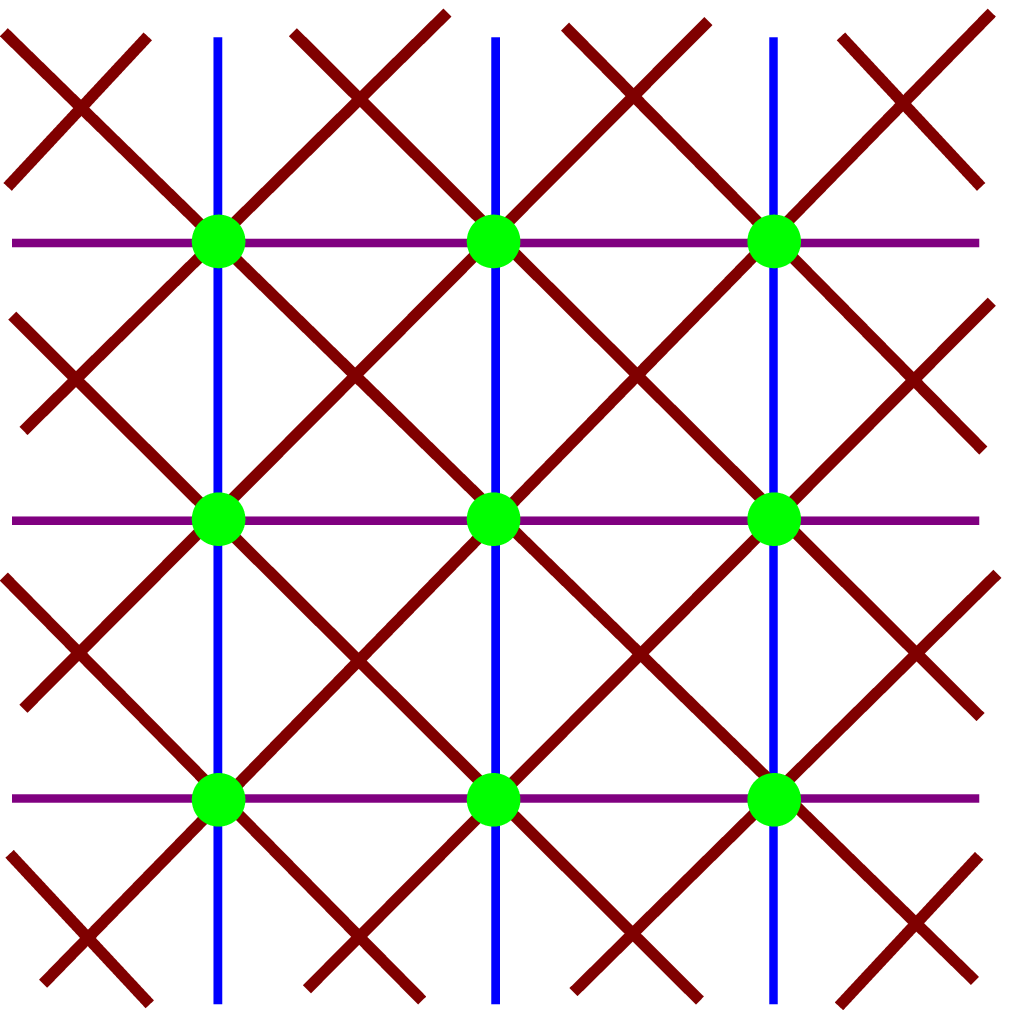}}\quad
\subfigure[\label{fig:NNN}]{\includegraphics[width=0.28\columnwidth]{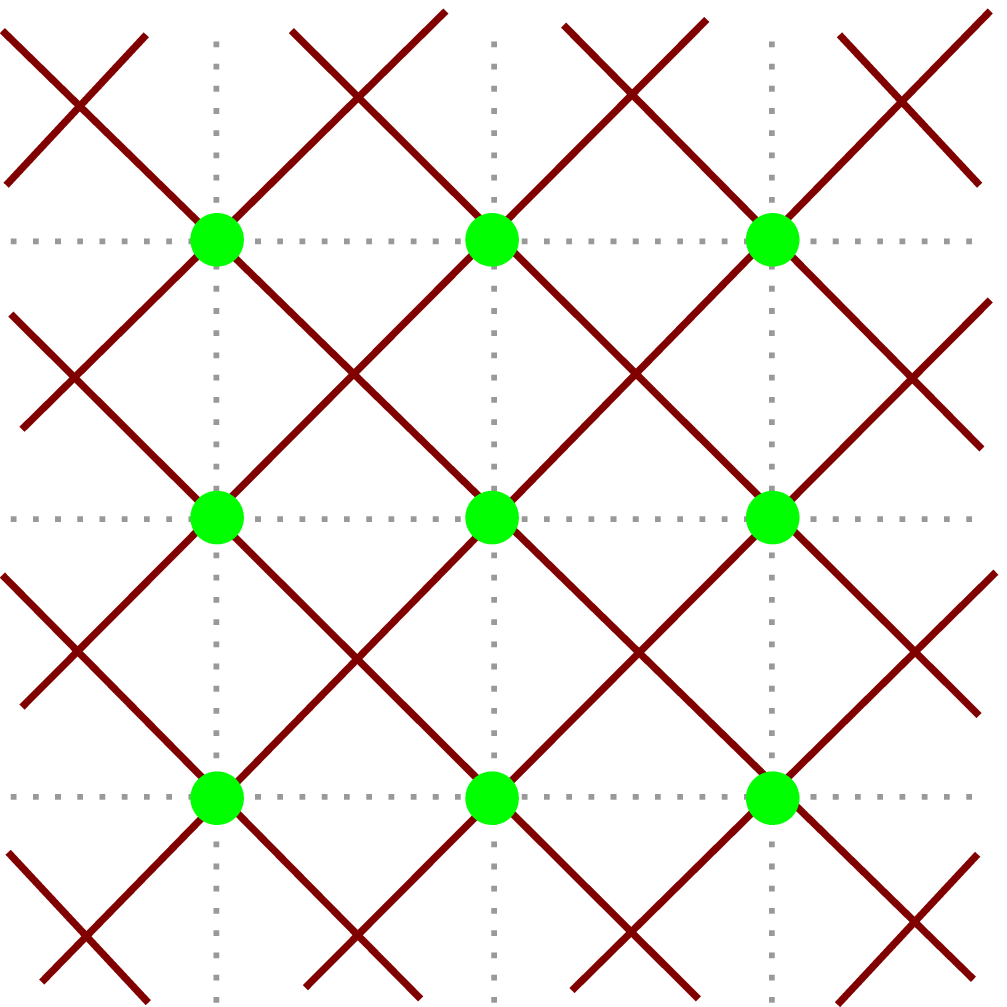}}\quad
\subfigure[\label{fig:Isolated}]{\includegraphics[width=0.28\columnwidth]{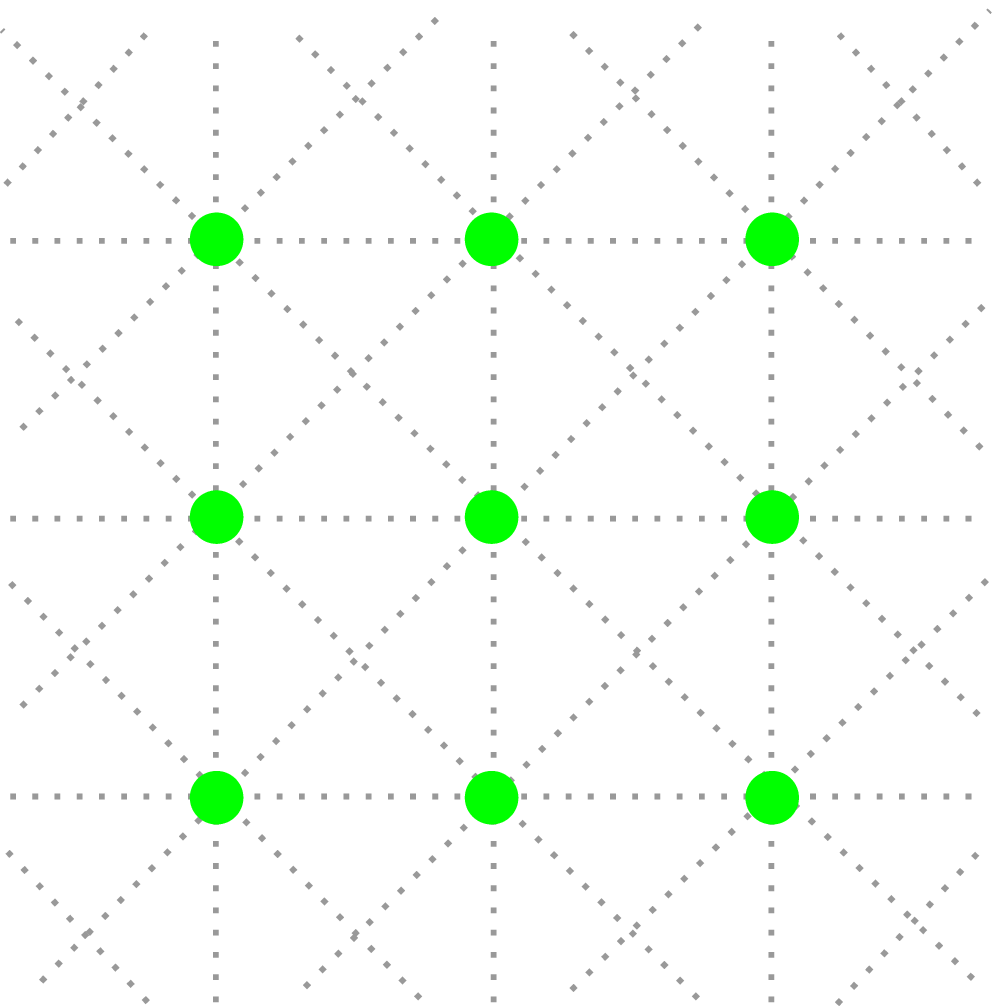}}
\caption{\label{fig:decoupling}\colorfig The real space picture of the renormalized hopping Hamiltonian on (a) the left of the ``NN decoupling'' line, (b) between the ``NN decoupling'' line and the ``NNN decoupling'' line, and (c) the right of the ``NNN decoupling'' line in Fig.~\ref{fig:phases}. Here the green dots denote the location of Fe atom, the violet (blue) lines denote non-zero renormalized nearest-neighbor hopping along the $x$ ($y$) direction, the brown lines denote non-zero renormalized next-nearest-neighbor hopping, and the gray dotted lines denote zero renormalized hopping.}%
\end{center}%
\end{figure}

As $J$ increases to $0.7 \lesssim J \lesssim 0.9$ , the onset of magnetism begins to shift to the left of the Mott transition. Moreover, the first-order boundary in which the nearest-neighbor bonds renormalize to zero starts to coincide with the Mott transition. The combined transition retains a first-order character, and the staggered magnetization exhibits a discontinuity across the transition. Furthermore, another phase boundary in which the \emph{next}-nearest-neighbor bonds also renormalize to zero (labeled as ``NNN decoupling'' in Fig.~\ref{fig:phases}) can now be seen, whose critical value $U_c^{\txt{NNN}}$ decreases with increasing $J$. Note that the phase boundary of this next-nearest-neighbor decoupling traces well with the onset of a fully polarized insulating state. This situation is typified in Fig.~\ref{fig:J=0.8}. For clarity, we also illustrate the real-space picture of the nearest-neighbor and next-nearest-neighbor decoupled state in Fig.~\ref{fig:decoupling}.

As $J$ further increases to $1.0 \lesssim J \lesssim 2.0$, the Mott transition becomes even stronger in first-order character and starts to coincide with the transition in which the next-nearest-neighbor bonds decouple. The phase transition is now accompanied by a jump to full polarization across the phase boundary. This situation is typified in Fig.~\ref{fig:J=1.2}. However, as $J$ further increases to $J \gtrsim 2.1$, the first-order character of the Mott transition starts to soften up, and, as typified by Fig.~\ref{fig:J=2.2}, a small shoulder in which the staggered magnetization is enhanced can now be observed immediately to the left of the Mott transition. 

Moreover, as can be seen in the inset of Fig.~\ref{fig:J=2.2}, an anisotropy of bond renormalization factors between the $x$ and $y$ nearest-neighbor bonds has also become discernible in this region where staggered magnetization is enhanced. In contrast, such anisotropy remains insignificant in all other parts of the parameter space we considered, including regions in the metallic phase in which the staggered magnetization is not enhanced. Note also that the bond renormalization factor is larger in the antiferromagnetic $x$ direction than in the ferromagnetic $y$ direction.

\section{Discussions and Conclusions} \label{sect:discussions}

\begin{figure}[tb]%
\begin{center}%
\subfigure[\label{fig:sym}]{\includegraphics[width=0.28\columnwidth]{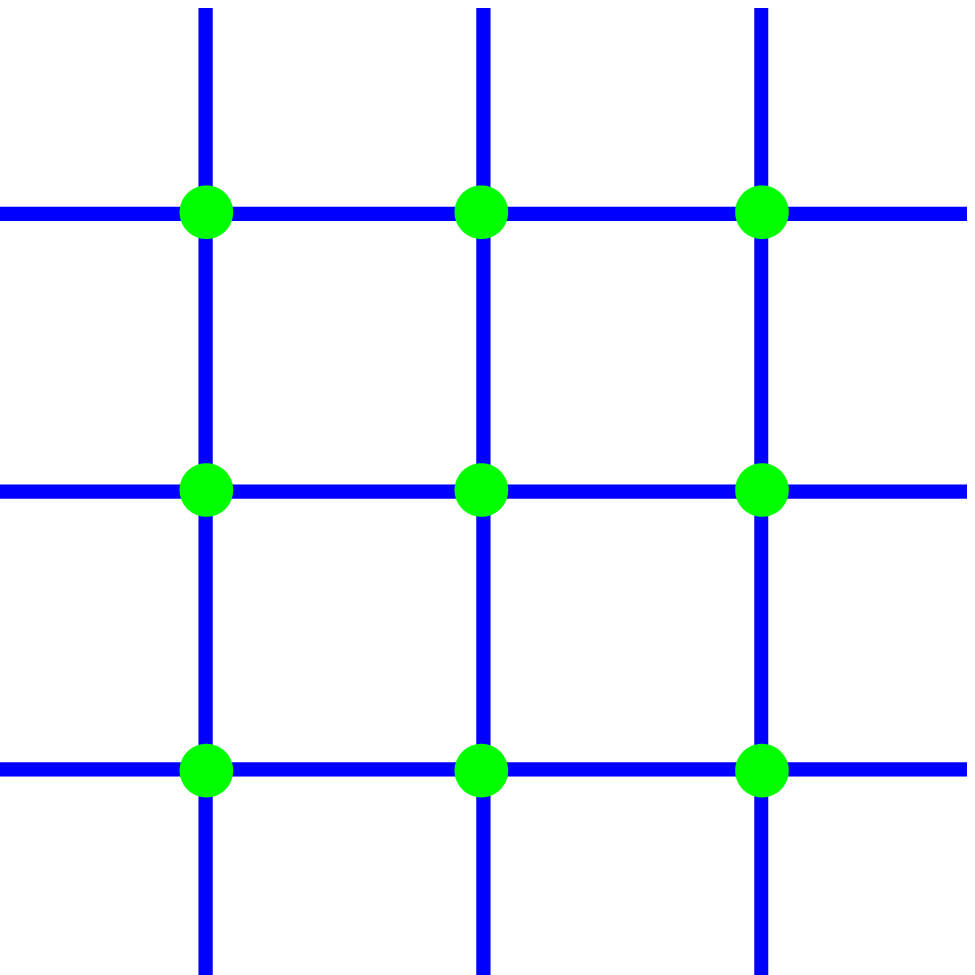}}\quad
\subfigure[\label{fig:symbroken}]{\includegraphics[width=0.28\columnwidth]{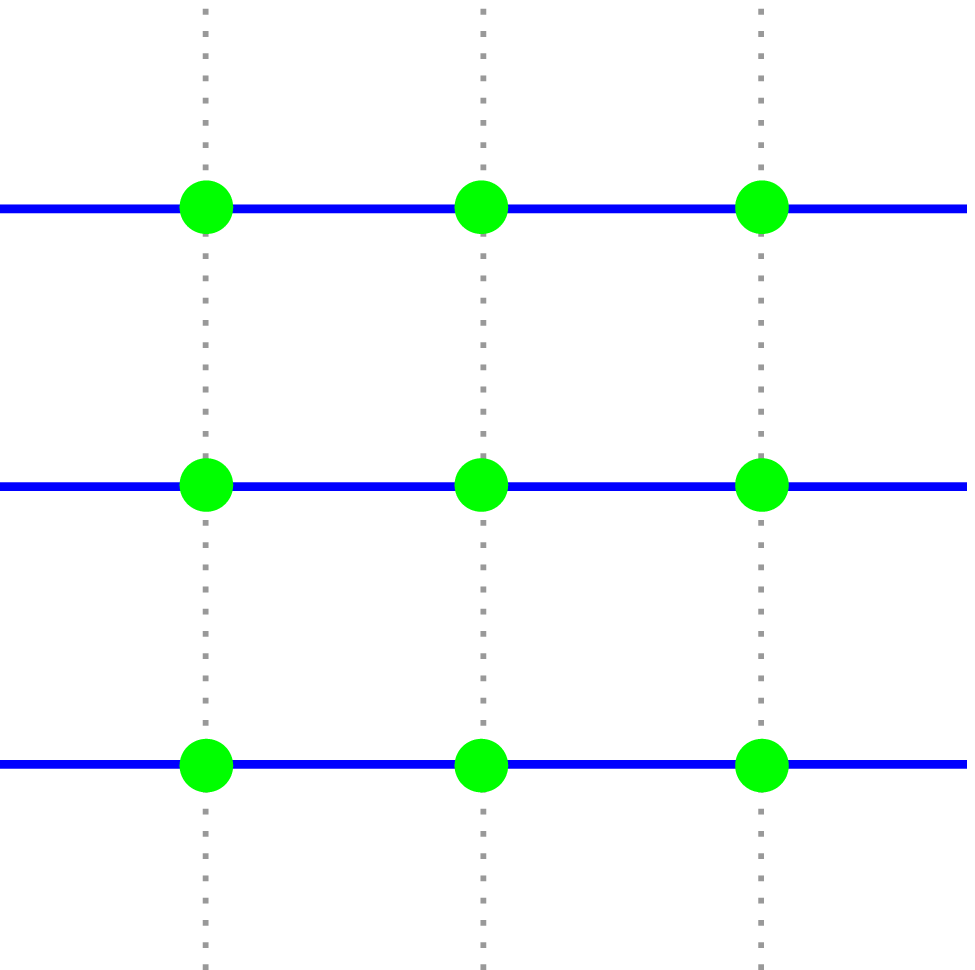}}\quad
\subfigure[\label{fig:twosites}]{\includegraphics[width=0.28\columnwidth]{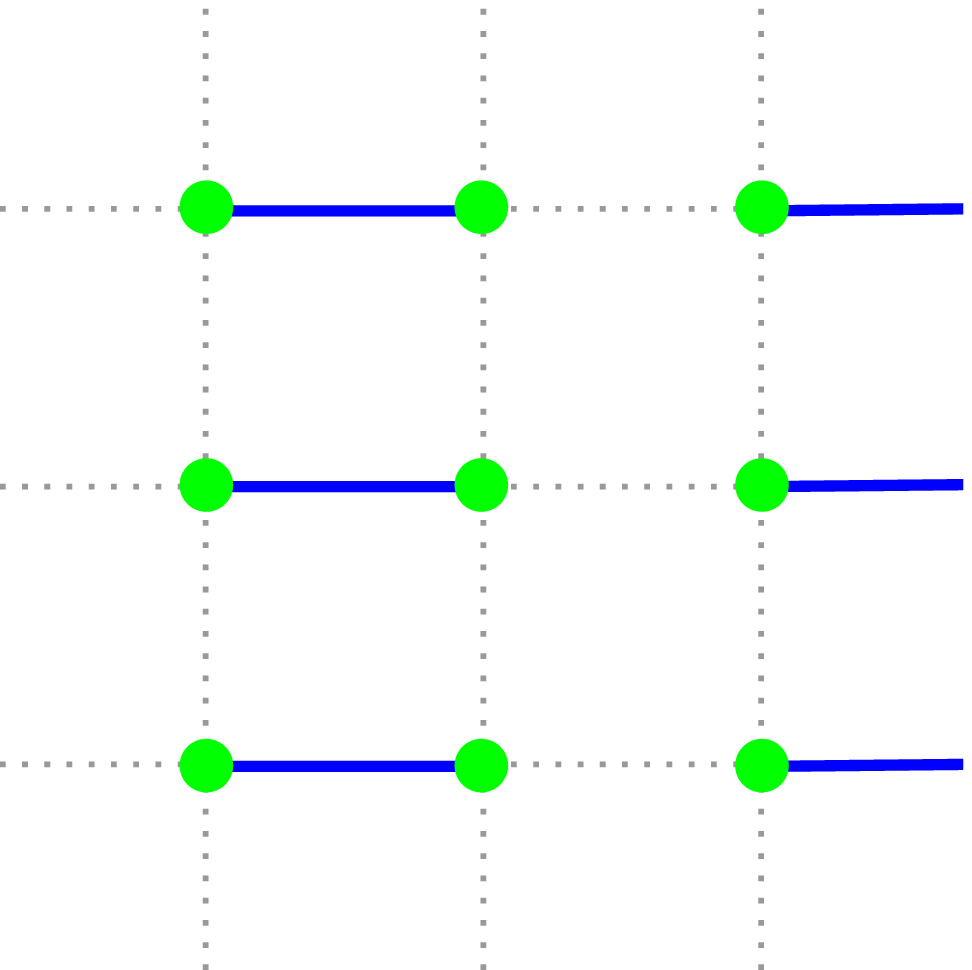}}
\caption{\label{fig:Sq}\colorfig Three patterns of renormalized bonds in a square lattice Hubbard model with non-mixing bands having only nearest-neighbor hopping. Here the green dots denote lattice sites, the blue line denote nonzero renormalized hopping, and the gray dotted lines denote zero renormalized hopping. It can be checked that within the slave-rotor mean field and on the insulating side (c) is energetically favorable to (b), which in turns is energetically favorable to (a).}%
\end{center}%
\end{figure}

Since the results in Sec.~\ref{sect:results} rely on various mean-field approximations, one should be cautious about the results thus obtained. In particular:
\begin{enumerate}
\item The critical value of $U$ for the Mott transition, $\Uc$, obtained in the slave-rotor approach is smaller than that obtained in the slave-spin approach.\cite{Yu:arXiv:1006.2337} This is a known discrepancy between the two methods and should vanish in the large-$N$ (number of orbitals) limit. For the case where $N$ is finite the slave-spin estimate is believed to be more accurate.\cite{deMedici:PRB:2005} This, however, should not affect the qualitative statements made in this paper.

\item In the slave-rotor formalism, all bonds that connect between the same sites are renormalized by the same factor, regardless of orbital character. This may lead to worries that some orbital selective aspects\cite{Werner:PRL:2007,Werner:PRB:2009} of the system are neglected. Such worries are partially alleviated by the study\cite{Yu:arXiv:1006.2337} of a similar two-band system in the slave-spin formalism, which allows for the possibility of spin-and-orbital dependent bond renormalizations and finds no evidence for an orbital selective state for $J > 0$. Furthermore, we reiterate that the possibility of an orbital symmetry broken SDW state is included in our formulation via an orbital-dependent mean-field parameter $M_{ab}$, but that we always find the staggered magnetization of the resulting state to be dominated by the orbital-diagonal component. Therefore, while orbital-selective scenarios have been proposed in the context of FeSCs,\cite{Kruger:PRB:2009, Hackl:NJP:2009, Laad:arXiv:1010.2940} we have found no supporting evidence for such state in the two-band model we considered.

\item As explained in Sec.~\ref{sect:formulation}, the distinct treatment for the Coulomb interaction and Hund's coupling is justified by the assumption that $U$ is the dominant scale. Generally, spin fluctuation is believed to play a more important role when the inter-orbital Coulomb interaction $U' = U - 2J$ is comparable to $J$, i.e., when $U < 3J$. Since $\Uc$ decreases rapidly as $J$ increases, the phase transition starts to enter the region where $U < 3J$ around $J \approx 2.5$. Therefore, the upper part of the phase diagram in Fig.~\ref{fig:phases} may be stretching the limit of validity for the present approach.

\item By treating $J$ via a weak-coupling approach, we have neglected the strong-coupling aspect of the Hund's coupling. In particular, one should not expect the $\Uc$ to remain unchanged for small $J$ as in Fig.~\ref{fig:phases}. Instead, one should expect $\Uc$ to drop immediately as $J$ increase, as is found\cite{Yu:arXiv:1006.2337} in the slave-spin approach.

\item In similar spirit, the present approximation has neglected the weak-coupling aspects of $U$. In particular, at the perturbative level both $U$ and $J$ are expected to drive the Fermi surface nesting at $(\pi,0)$. Hence, the paramagnetic metal and the spin-liquid region in Fig.~\ref{fig:phases} should, if exist at all, be greatly reduced in size.

\item One peculiar feature of the mean-field result presented in Sec.~\ref{sect:results} is the existence of phases in which certain bonds in the hopping Hamiltonian is renormalized to zero in the insulating phase. This seems to be a general feature of the slave-rotor method, since similar results is also found in a Hubbard model with a different lattice geometry (and which the Hund's coupling is dropped).\cite{Yu:arXiv:1101.3307} Indeed, the problem can already be seen in the simplest case of a square-lattice Hubbard model with non-mixing bands having only nearest-neighbor hoppings. In such case it can be checked that on the insulating side, within the slave-rotor mean field, a symmetry broken phase in which all bonds along a particular direction is renormalized to zero is energetically favorable to a state in which all bonds are renormalized by the same amount. This rotation-symmetry broken state is, in turn, less energetically favorable to a translation-and-rotation-symmetry broken state in which the bonds renormalize to a dimer pattern (see Fig.~\ref{fig:Sq} for illustration). Since the dimer pattern is known to be favorable in the large-$N$ limit,\cite{Marston:PRB:1989,Read:NPB:1989} the decoupled phase in Fig.~\ref{fig:phases} may be an artifact of the large-$N$ approximation.
\end{enumerate}

In spite of the issues mentioned above, the present study may still shed light on the question we posed in the introduction of this paper: namely, whether there are any signs of local-moment physics on the metallic side of the phase diagram. For this, observe that $\Uc$ remains in the region where $U > 3J$  for $J$ in the range $0.7 \lesssim J \lesssim 2.5$, and as we approach the phase boundary from $U < \Uc$ the bond renormalization remains modest for much of the parameter space, such that the renormalized bandwidth $\bar{W}$ is at least comparable to $J$. In such region the approximation we took in this paper may be valid. And for $0.7 \lesssim J \lesssim 2.1$, as we approach the phase boundary from $U < \Uc$ there is no sign of magnetization enhancement (which, in this formulation, will be accompanied by a sharper-than-usual drop in $Q^{f}_{ij}$) until the first-order phase boundary is hit. Thus, our result is suggestive that in this range of $J$ there is no sign of local-moment physics on the metallic side. In contrast, for $2.1 \lesssim J \lesssim 2.5$ there is a sharper drop in $Q^{f}_{ij}$ as we approach the phase boundary, which suggests the possibility that in this range of $J$ some remnant of local-moment physics can be found.

\begin{acknowledgments}
We thank Fa Wang, Ying Ran, and Senthil Todadri for helpful discussions. This research is supported in part by DOE under Grant No.\@ DE-FG02-03ER46076 (WHK and PAL) and in part by NSF under Grant No.\@ PHY05-51164 (WHK).
\end{acknowledgments}

%%%% Use existing bibliography file

%%%% Generate bibliography from bibtex
%\bibliographystyle{apsrev}
%\bibliography{../../../Papers/references} 

\end{document}